\documentclass[%
 reprint,
superscriptaddress,
 amsmath,amssymb,
 aps, prl,
]{revtex4-1}

\usepackage{graphicx} 
\usepackage{dcolumn} 
\usepackage{sidecap}
\usepackage{bm} 
\usepackage[svgnames]{xcolor}
\usepackage[bookmarks=false,linkcolor=blue,urlcolor=blue,colorlinks,citecolor=blue]{hyperref}

\begin{document}

\title{Local and Nonlocal Transport Spectroscopy in Planar Josephson Junctions}

\author{A.~Banerjee}
\affiliation{Center for Quantum Devices, Niels Bohr Institute,
University of Copenhagen,
Universitetsparken 5, 2100 Copenhagen, Denmark
}

\author{O.~Lesser}
\affiliation{Department of Condensed Matter Physics, Weizmann Institute of Science, Rehovot, Israel 76100
}

\author{M.~A.~Rahman}
\affiliation{Center for Quantum Devices, Niels Bohr Institute,
University of Copenhagen,
Universitetsparken 5, 2100 Copenhagen, Denmark
}

\author{C.~Thomas}
\affiliation{Department of Physics and Astronomy, and Birck Nanotechnology Center,
Purdue University, West Lafayette, Indiana 47907 USA
}

\author{T.~Wang}
\affiliation{Department of Physics and Astronomy, and Birck Nanotechnology Center,
Purdue University, West Lafayette, Indiana 47907 USA
}

\author{M.~J.~Manfra}
\affiliation{Department of Physics and Astronomy, and Birck Nanotechnology Center,
Purdue University, West Lafayette, Indiana 47907 USA
}

\affiliation{School of Materials Engineering, and School of Electrical and Computer Engineering, Purdue University, West Lafayette, Indiana 47907 USA}

\author{E.~Berg}
\affiliation{Department of Condensed Matter Physics, Weizmann Institute of Science, Rehovot, Israel 76100
}

\author{Y.~Oreg}
\affiliation{Department of Condensed Matter Physics, Weizmann Institute of Science, Rehovot, Israel 76100
}

\author{Ady Stern}
\affiliation{Department of Condensed Matter Physics, Weizmann Institute of Science, Rehovot, Israel 76100
}

\author{C.~M.~Marcus}
\affiliation{Center for Quantum Devices, Niels Bohr Institute,
University of Copenhagen,
Universitetsparken 5, 2100 Copenhagen, Denmark
}


\begin{abstract}
We report simultaneously acquired local and nonlocal transport spectroscopy in a phase-biased planar Josephson junction based on an epitaxial InAs/Al hybrid two-dimensional heterostructure. Quantum point contacts at the junction ends allow measurement of the 2$\times$2  matrix of local and nonlocal tunneling conductances as a function of magnetic field along the junction, phase difference across the junction, and carrier density. A closing and reopening of a gap was observed in both the local and nonlocal tunneling spectra as a function of magnetic field. For particular tunings of junction density, gap reopenings were accompanied by zero-bias conductance peaks (ZBCPs) in local conductances. End-to-end correlation of gap reopening was strong, while correlation of local ZBCPs was weak. A simple, disorder-free model of the device shows comparable conductance matrix behavior associated with a topological phase transition. Phase dependence helps distinguish possible origins of the ZBCPs.
\end{abstract}

\maketitle
Topological materials obey a bulk-boundary correspondence, establishing a connection between the topological index of the bulk and the number of boundary modes~\cite{QiZhang,FuKane}. In one-dimensional topological superconductors (1D-TSCs)~\cite{Oreg,Lutchyn}, these considerations imply that the bulk modes undergo a characteristic closing and reopening of the superconducting gap whenever the system is driven through a topological phase transition. In this situation, the gap reopening is connected to the appearance of zero-energy Majorana modes at the two ends~\cite{SauReopening,StanescuReopening,TewariReopening,RainisReopening}. Several experimental works have reported zero-bias conductance peaks (ZBCPs)  at the ends of wires or 1D structures, identified as signatures of 1D-TSCs~\cite{Mourik, NadjPerge, Das, Deng, NicheleScaling,grivnin, fornieri}. However, in most of these cases, though not all \cite{Vaitiekenas, GapReopening}, an associated gap closing and reopening was not observed in tunneling conductance. 

An emerging method that allows simultaneous observation of end modes and bulk gap behavior is nonlocal spectroscopy, where measurement of the tunneling current between the ends of a device provides information about the bulk~\cite{Akhmerov}. This technique requires a three-terminal (3T) configuration, and has been theoretically explored for nanowires~\cite{Akhmerov,Sarma3T,DanonNonlocal,Hess2021local,wang2021non}. Nonlocal transport experiments, also in nanowires, were used to probe symmetries of the conductance matrix arising from current conservation and measure the local charge of Andreev bound states~\cite{MenardNonlocal, DanonNonlocal}. Experiments in short nanowire segments identified end-to-end correlation between extended Andreev bound states~\cite{AnselmettiNonLocal}. In long nanowire segments, local conductance showed ZBCPs while the gap in the nonlocal spectrum remained closed~\cite{PugliaNonlocal}, suggesting non-topological ZBCPs arising from strong disorder~\cite{AltlandZBCP, PradaZBCP,LiuZBCP,DasSharmaZBCP,DasSharmaZBCP2}. These experiments demonstrated the importance of combining local and nonlocal transport to differentiate  trivial and potentially topological ZBCPs.

\begin{figure}[t]
\includegraphics[width=0.5\textwidth]{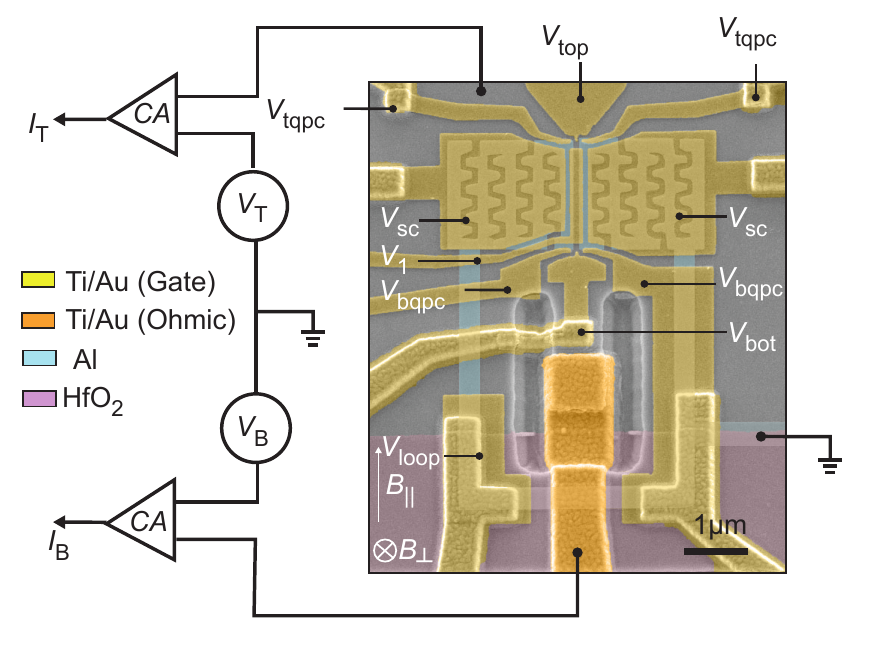}
\caption{\label{fig01} {\bf Device and measurement setup.}~False-color micrograph of a representative device showing three-terminal configuration. Meandering perforations etched onto the superconducting leads allow partial depletion of the semiconductor using gate voltage $V_{\rm sc}$. Al loop allows phase biasing of the junction with a small out-of-plane magnetic field, $B_{\perp}$. An in-plane magnetic field, $B_{\parallel}$ is applied parallel to the S-N interfaces. Voltage biases $V_{\rm T}$ and $V_{\rm B}$ are applied to the top and bottom ohmic contacts through the current amplifiers (CA). Gates $V_{\rm top(bot)}$ and $V_{\rm t(b)qpc}$ form QPCs at the junction ends. $V_{1}$ controls carrier density in the junction. All connections to the device are via $\sim 1-2$k$\Omega$ fridge wires and filters, see Supplemental Material for details.}
\end{figure}

Planar Josephson junctions (PJJs) of superconductor-semiconductor hybrids have recently emerged as a promising alternative platform for topological superconductivity, providing several knobs that can control a possible topological superconducting phase, including, notably, the novel control parameter of the phase difference across the junction~\cite{HellFlensberg,Pientka,Setiawan,fornieri,ShabaniReopening,ren,GoswamiReopening,Nichele2022}. 

In this Letter, we extend previous work \cite{ GapReopening} by investigating nonlocal conductance spectroscopy, measured simultaneously with local spectroscopy, on a 3T PJJ device with quantum point contacts (QPCs) at both ends, measured  as a function of in-plane magnetic field along the junction, phase difference across the junction, and carrier density within the junction. Our main observation is a closing and reopening of the superconducting gap in {\it nonlocal} conductance correlated with the appearance of ZBCPs in {\it local} conductance. We find that the gap closing and reopening in both local and nonlocal conductance is robust against variations of the junction carrier density, but the observation of ZBCPs at one or both ends require careful tuning of voltages on the junction and QPC gates. 

To help interpret these results, we investigate numerically a simple model of a PJJ~\cite{GapReopening}. Within the model, a gap reopening in nonlocal conductances (without ZBCP) appearing together with a ZBCP in local conductance is characteristic of a topological phase transition. 

Figure~\ref{fig01} shows a micrograph of one of the devices, along with a schematic electrical circuit. The PJJ can be probed by a pair of integrated QPCs at the ends of the junction, and phase-biased by applying a small ($\sim 0.1$~mT scale) out-of-plane magnetic field through a superconducting loop. 

The device was fabricated on a molecular-beam-epitaxy grown heterostructure stack with a shallow InAs quantum well separated from a top Al layer by an In$_{0.75}$Ga$_{0.25}$As barrier. A combination of wet etching of the Al layer and deep wet etching of the semiconductor stack was used to define the superconducting loop, the Josephson junction and the mesa with a U-shaped trench. A patch of the mesa (with Al removed) within the loop was contacted by a layer of Ti/Au to form an internal submicron ohmic contact to enable bottom-end tunneling spectroscopy. A layer of HfO$_2$, grown by atomic layer deposition (ALD) and patterned in a rectangular shape, was used to isolate the Ti/Au layer from the superconducting loop and the conducting mesa. A second layer of HfO$_2$ was deposited globally followed by the deposition of Ti/Au gates for electrostatic control of the junction and the QPCs. 

The carrier density in the normal barrier of the JJ (width $w_n=100$~nm, length $l=1.6\,\mu$m) was controlled by gate voltage $V_1$. Gate voltage $V_{\rm sc}$ controlled the carrier density in the semiconductor underneath the superconducting leads. Split gates controlled by voltages $V_{\rm tqpc}$ and $V_{\rm bqpc}$ define QPCs at the top and bottom of the junction. Additional gate voltages $V_{\rm top}$ and $V_{\rm bot}$ controlled densities in the normal regions outside the QPCs, and were typically fixed at $\sim +100$~mV.

\begin{figure}[t]
\includegraphics[width=0.5\textwidth]{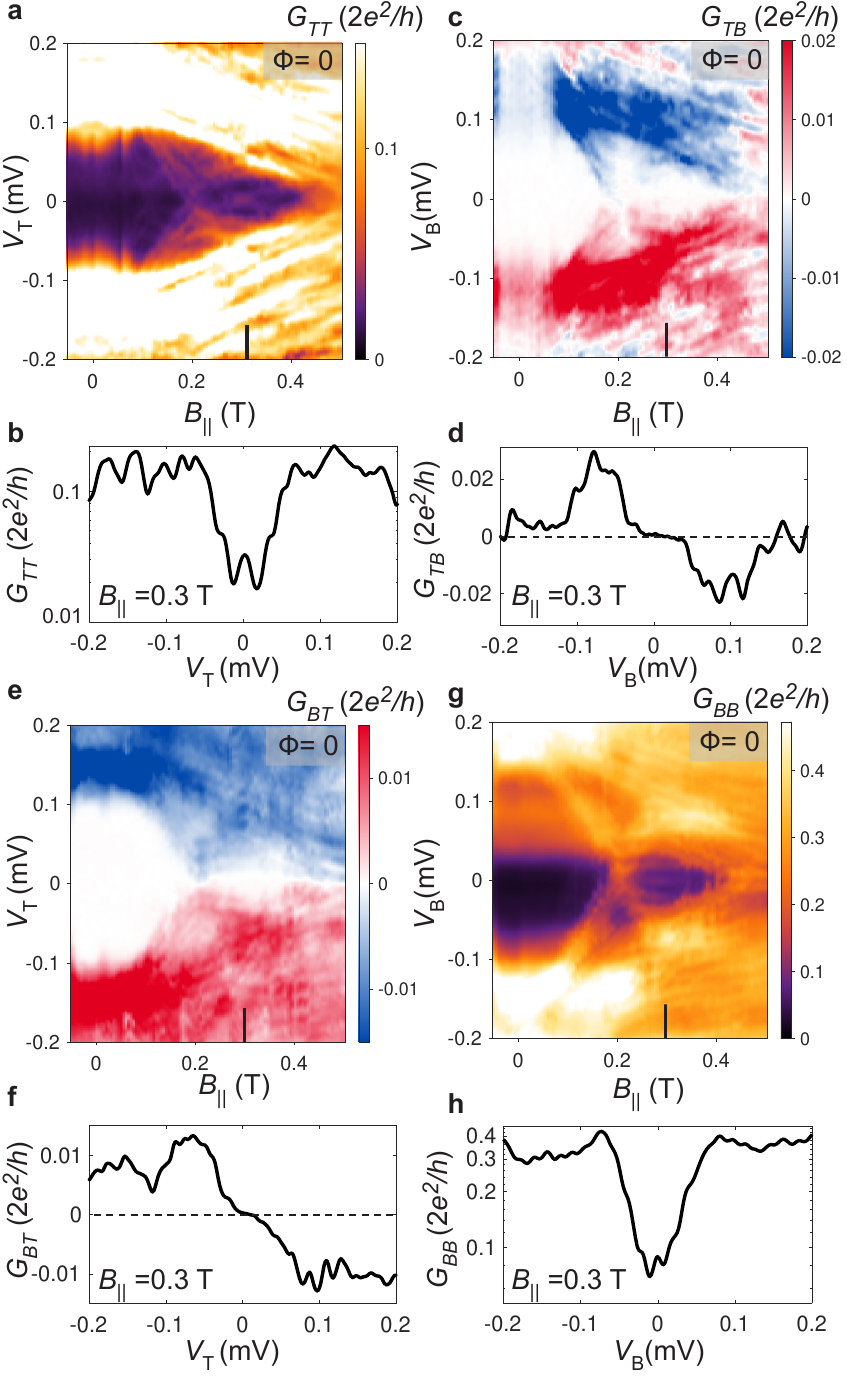}
\caption{\label{fig02} {\bf Magnetic field dependence of the conductance matrix.}~(a)~Local differential conductance $G_{\rm TT}$ and (e)~Nonlocal differential conductance $G_{\rm BT}$ measured as a function of $V_{\rm T}$ and $B_{\parallel}$. (c)~Nonlocal differential conductance $G_{\rm TB}$ and (g)~local differential conductance $G_{\rm BB}$ measured as a function of $V_{\rm B}$ and $B_\parallel$. The phase bias is set to $\Phi=0$. Line-cuts at $B_\parallel=0.3$~T where (b)~$G_{\rm TT}$ shows a ZBCP. (h)~$G_{\rm BB}$ shows a ZBCP. (d)~$G_{\rm TB}$ and (h)~$G_{\rm BT}$ are strongly antisymmetric at high DC biases, and zero in a finite range around zero DC bias. Gate voltage settings used for this measurement were $V_{\rm tqpc}=-0.37$~V, $V_{\rm bqpc}=-0.34$~V, $V_1=0.085$~V and $V_{\rm sc}=-3.6$~V.}
\end{figure}

\begin{figure}[t]
\includegraphics[width=0.5\textwidth]{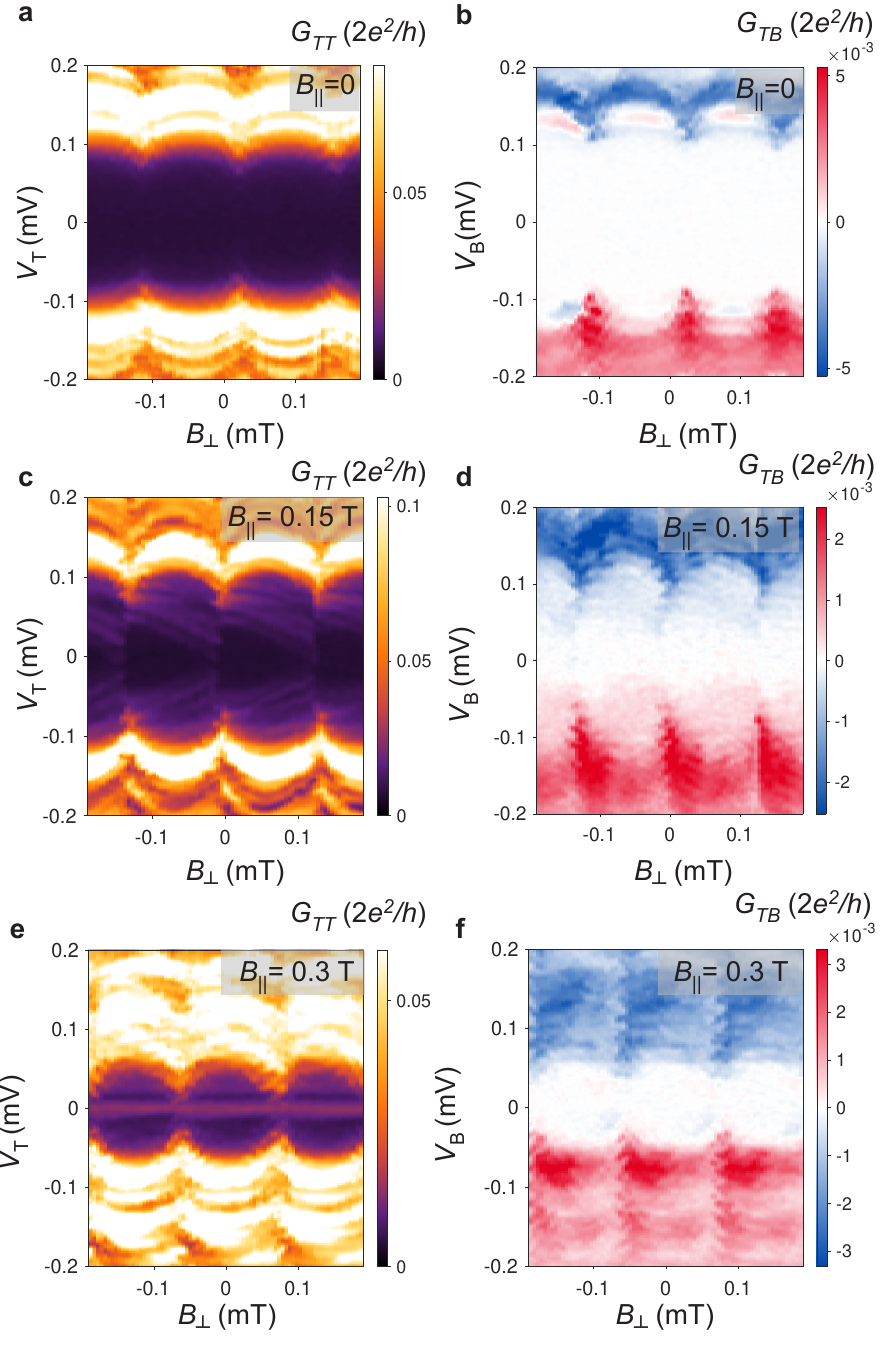}
\caption{\label{fig03} {\bf Phase dependence of local and nonlocal conductance spectra.} Local differential conductance $G_{\rm TT}$ (left column) and nonlocal differential conductance $G_{\rm TB}$ (right column) measured as a function of out-of-plane magnetic field $B_\perp$ and source-drain bias, $V_{\rm T}$ and $V_{\rm B}$, respectively at three values of the in-plane magnetic field. (a) and (b)~$B_\parallel=0$, showing a periodic modulation of the superconducting gap in both the local and nonlocal spectrum. The nonlocal spectrum shows a larger amplitude of the superconducting gap than the local conductance spectrum. (c) and (d)~$B_\parallel=0.15$~T, shows sub-gap states that are lowered in energy. These states are phase-asymmetric and appear in both $G_{\rm TT}$ and $G_{\rm TB}$. (e) and (f)~$B_\parallel=0.3$~T shows a phase-independent ZBCP in $G_{\rm BB}$, but not in $G_{\rm TB}$. $G_{\rm TB}$ displays a superconducting gap that is modulated periodically with $B_\perp$. Gate voltage settings used for this measurement were $V_{\rm tqpc}=-0.385$~V, $V_{\rm bqpc}=-0.38$~V, $V_1=0.092$~V and $V_{\rm sc}=-3.6$~V.}
\end{figure}

The 3T measurement configuration is shown in Fig.~\ref{fig01}(a). The top ohmic contact is a region of InAs separated from the junction by the top QPC.  The bottom ohmic contact is formed by a Ti/Au electrode, separated by the bottom QPC. The Al loop connecting the two sides of the junction provides the third contact, held at ground. Low-frequency AC plus DC voltage biases $V_{\rm T(B)}$ are applied through current amplifiers (denoted CA). The measured currents  $I_{\rm T(B)}$ then yield the 2$\times$2 conductance matrix, $G_{\rm ij}=dI_{\rm i}/dV_{\rm j}$, with ${\rm i},{\rm j}=\rm{T,B}$ via standard AC lock-in measurements (see additional details in Supplemental Material).

For measurements shown in Fig.~\ref{fig02}, the conductance matrix was measured as a function of in-plane magnetic field, $B_\parallel$, with $V_{\rm sc}=-3.6$~V, giving a hard superconducting gap in the leads, $V_1=+85$~mV, giving ZBCPs in both top and bottom local conductances at $B_\parallel \sim 0.3$~T, and QPCs set to $V_{\rm tqpc}=-0.37$~V, $V_{\rm bqpc}=-0.34$~V, to yield sizable nonlocal conductances, $\sim 0.01 \times 2e^2/h$ near the gap edge, $|V_{\rm T,B}|\sim150~\mu$V. To compensate any coupling of $B_\parallel$ through the superconducting loop controlling phase $\Phi$ across the junction, a sweep of $B_\perp$ was made at each value of $B_\parallel$ and then sliced along cuts of constant $\Phi$ numerically by following $\Phi$-dependent lobe features (see Methods). This allowed us to obtain the $B_{\parallel}$ dependence of the conductance matrix at fixed flux, as shown, for instance, in  Fig.~\ref{fig02} for $\Phi=0$. 

Local conductance spectra showed a finite superconducting gap around $B_\parallel=0$ [Fig.~\ref{fig02}(a, g)]. With increasing $B_\parallel$, a band of resolvable discrete states moved towards zero bias, closing the gap at $B_\parallel \sim$~0.2~T followed by its reopening. Beyond the reopening  (0.2~T~$<B_\parallel<0.4$~T), but not before, ZBCPs were observed in both $G_{\rm TT}$ and $G_{\rm BB}$ [Fig.~\ref{fig02}(b, h)]. In this data set, the ZBCP at the top end splits as $B_\parallel$ approaches 0.4~T, whereas the bottom end ZBCP appears to remain at zero, but diminishes in amplitude. Additionally, the ZBCPs observed at each end do not exhibit strong correlation with respect to variations of $V_1$ (see Fig.~\ref{suppfigS9}). Both local conductances show a final gap closure at $B_\parallel \sim~0.45$~T.

The corresponding behavior of the nonlocal conductance spectra is shown in Figs.~\ref{fig02}(c, e). A predominantly antisymmetric signal is observed throughout the measured magnetic field range, with amplitude remaining roughly uniform. The gap in the nonlocal spectrum undergoes a closure at $B_\parallel \sim 0.2$~T, at the same magnetic field as the local conductance spectrum, and is visible in both $G_{\rm TB}$ and $G_{\rm BT}$. Both nonlocal conductances remain strongly antisymmetric around zero bias. The nonlocal gap then reopens obtaining a maximum at $B_\parallel \sim$ 0.3~T, with line-cuts shown in Figs.~\ref{fig02}(d, f). Notably, no ZBCP is observed. Both nonlocal conductances disappear in a finite window around zero bias before turning on sharply at finite V$_{\rm T/B}$. The final closure of the nonlocal gap at $B_\parallel \sim $~0.45~T is more pronounced, in terms of signal strength, than the closure at $B_\parallel \sim $~0.2~T.

Local and nonlocal conductance spectra are modulated by a small perpendicular magnetic field, $B_\perp$, which threads flux through the $\sim 12\,\mu$m$^2$ superconducting loop ($B_\perp \sim ~$0.17~mT corresponds to $\Phi_0=h/2e$ through the loop), showing the same period in $B_\perp$ and in phase. Around $B_\parallel=0$ [Figs.~\ref{fig03}(a, b)], the local gap in $G_{\rm TT}$ appears smaller than the nonlocal gap in $G_{\rm BT}$. At $B_\parallel \sim 0.15$~T~[Figs.~\ref{fig03}(c),(d)], flux-dependent states are lowered in energy and fill the sub-gap spectrum in both local and nonlocal conductances. Within each flux lobe, states are asymmetric with respect to $\Phi$. At $B_\parallel=$0.3~T, a phase-independent ZBCP is measured in $G_{\rm TT}$ [Fig.~\ref{fig03}(e)] but absent in $G_{\rm BT}$ [Fig.~\ref{fig03}(f)]. At this field, $G_{\rm BT}$ remains zero until a source-drain bias of $V_{\rm T}\sim40~\mu$eV at $\Phi=0$ and closes at $\Phi=\Phi_0/2$. $G_{\rm TB}$ and $G_{\rm BB}$, are qualitatively similar to $G_{\rm BT}$ and $G_{\rm TT}$ respectively (see Fig.~\ref{suppfigS8}).

\begin{figure}[t]
\includegraphics[width=0.5\textwidth]{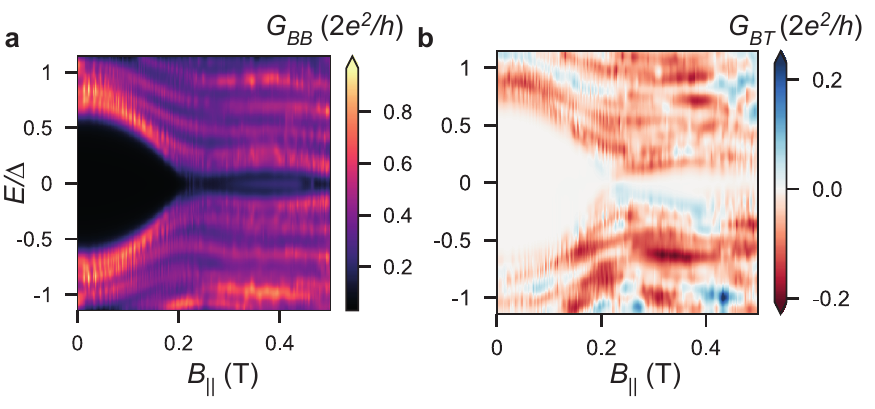}
\caption{\label{fig04} {\bf Numerical simulation of local and nonlocal conductance.} (a)~Local conductance $G_{\rm BB}$ evaluated at the bottom end of the junction as a function of $B_\parallel$, and $\Phi=$~0 including thermal broadening equivalent to a temperature of 50~mK. At $B_\parallel \sim 0.2$~T the junction undergoes a topological phase transition reflected by the closing and reopening of the superconducting gap, followed by the appearance of a ZBCP. (b)~Nonlocal conductance $G_{\rm BT}$ shows a corresponding gap-reopening transition at $B_\parallel \sim 0.2$~T without the formation of a ZBCP.}
\end{figure}

We also investigated nonlocal transport at gate settings where a ZBCP was observed in the bottom local conductance, but not the top local conductance (see Fig.~\ref{suppfigS3} and \ref{suppfigS4}). In this case, a gap-reopening signature was observed in nonlocal conductance and the nonlocal gap remained finite in the reopened state. In other devices where reasonably strong nonlocal conductance was observed ($\sim 0.01\times2e^2/h$), the nonlocal spectrum exhibited a gap-reopening feature. In some cases, the sub-gap nonlocal conductance in the reopened state remained finite, indicating a soft nonlocal gap (see Figs.~\ref{suppfigS10} and ~\ref{suppfigS11}).

To help interpret characteristic features of the observed conductance matrix, we investigate a simple model of a PJJ using the Kwant software package~\cite{groth_kwant_2014}, as described previously~\cite{GapReopening}. Here, we extend the model by tunnel coupling the system to metallic leads at the junction ends. Top-bottom symmetry of the model ensures that $G_{\rm TT}=G_{\rm BB}$ and $G_{\rm TB}=G_{\rm BT}$. Figure~\ref{fig04}(a) shows the local conductance spectrum undergoing a topological gap-reopening transition at $B_\parallel \sim 0.2$~T, followed by a ZBCP arising from a Majorana zero mode. The corresponding nonlocal conductance spectrum, shown in Fig.~\ref{fig04}(b), also shows a reopening of the gap, but no sub-gap structure once the gap reopens. 

These model results support the interpretation that nonlocal conductance is mediated by a combination of co-tunneling and crossed-Andreev reflection of quasiparticles carried by extended Andreev bound states. These states have a finite probability density throughout the length of the junction, including the two ends, and therefore also appear in the local conductances. Within this picture, Majorana zero-modes appear as zero-bias peaks only in local conductance, not in nonlocal conductance because of their localized probability density. This is in contrast to extended Andreev bound states, which are expected to appear in both local and nonlocal conductances~\cite{MenardNonlocal, Hess2021local}.

Conductance matrix signatures obtained from our model are roughly consistent with experiment. However, as opposed to the experiment, in the numerical simulations we find a large symmetric component of the nonlocal conductance, comparable in strength to the antisymmetric component. Their relative strength depends on details used to model finite temperature, disorder, and tunneling barriers~\cite{Akhmerov,wang2021non}, and may explain this discrepancy. Finally, local ZBCPs are fully correlated in the model, but lack such correlation in experiment.

Within a non-topological interpretation of our data, an inhomogeneous chemical potential profile produces non-topological zero-energy Andreev bound states at the two device ends. In the model these states are not stable at zero-energy with respect to variation of $\Phi$, which may provide a distinguishing signature. Another non-topological scenario is the case of strong disorder, where a proliferation of low-energy sub-gap states prevents a topological phase transition. This scenario can produce ZBCPs in local conductance, but tends not to show a gap-reopening in the nonlocal conductance~\cite{Sarma3T,PugliaNonlocal}. 

Within a topological interpretation, the presence of a finite nonlocal gap without strong end-to-end ZBCP correlation may arise from disorder of weak-to-intermediate strength~\cite{woods2021charge}. Such disorder creates disjointed topological segments in the bulk of the junction, eliminating end-to-end ZBCP correlation while preserving the nonlocal gap together and gap-reopening signature. Further study of conductance matrix behavior with a greater degree of the potential along the junction would help clarify the situation.

We thank Geoff Gardner and Sergei Gronin for contributions to materials growth, and  Asbj{\o}rn Drachmann for assistance with fabrication. We thank Lucas Casparis, Tom Dvir, Karsten Flensberg, Max Geier, Esteban Martinez, and Andreas P\"{o}schl, for valuable discussions. We acknowledge a research grant (Project 43951) from VILLUM FONDEN, support from the ERC under the Horizon 2020 Research and Innovation programme (LEGOTOP No. 788715 and HQMAT No. 817799), the DFG (CRC/Transregio 183, EI 519/7-1), the BSF and NSF (2018643), the ISF Quantum Science and Technology (2074/19), and a research grant from Irving and Cherna Moskowitz.

\bibliography{Gap_reopening_3T.bib}

\clearpage

\onecolumngrid
\appendix
\begin{center}
{\bf Supplemental Material}
\end{center}

\setcounter{equation}{0}
\renewcommand{\theequation}{S\arabic{equation}}
\setcounter{figure}{0}
\renewcommand{\thefigure}{S\arabic{figure}}
\setcounter{section}{0}
\renewcommand{\thesection}{S\Roman{section}}

{\bf Wafer structure:} The heterostructure stack used in this work consists of an InAs two-dimensional quantum well in epitaxial contact with aluminum, grown by molecular beam epitaxy. The wafer was grown on an insulating InP substrate and comprises a 100-nm-thick In$_{0.52}$Al$_{0.48}$As matched buffer, a 1$\mu$m thick step-graded buffer realized with alloy steps from In$_{0.52}$Al$_{0.48}$As to In$_{0.89}$Al$_{0.11}$As (20 steps, 50 nm/step), a 58 nm In$_{0.82}$Al$_{0.18}$As layer, a 4 nm In$_{0.75}$Ga$_{0.25}$As bottom barrier, a 7 nm InAs quantum well, a 10 nm In$_{0.75}$Ga$_{0.25}$As top barrier, two monolayers of GaAs and a 7 nm film of epitaxially grown Al. The top Al layer was grown without breaking the vacuum, in the same growth chamber.  Hall effect measurements performed in Hall bar devices of the same material, with Al etched away, indicated a peak electron mobility peak $\mu= 43,000$~ cm$^2$/V-s at a carrier density of $n=8 \times 10^{11}$~cm$^{-2}$, corresponding to a peak electron mean free path of  $l_e\sim$~600 nm. This suggests that our devices are quasiballistic along the length $l \sim 3 l_e$ and ballistic in the width direction $w_n \sim l_e/6$. Transport characterization of an etched aluminum Hall bar revealed an upper critical field $B_\parallel \sim$~2.5~T, indicating that the Al layer remains superconducting well above the collapse of the induced superconducting gap at $B_\parallel \sim$~0.5~T.  

\vspace{10 pt}

{\bf Device fabrication:} Devices were fabricated using standard electron beam lithography. Devices on the same chip were electrically isolated from one another using a two-step mesa etch process, first by removing Al with Al etchant Transene D, and then a standard III-V chemical wet etch using a solution H$_2$O : C$_6$H$_8$O$_7$ : H$_3$PO$_4$ : H$_2$O$_2$ (220:55:3:3) to etch the mesa until a depth of $\sim$300 nm. This step also defined the U-shaped trench and the patch of mesa that eventually formed the submicron ohmic contact. In the next lithography step, the Al layer was selectively removed leaving behind the Josephson junction with a flux loop. Particularly, this step also removed Al from the internal ohmic mesa patch. A patterned layer of dielectric, comprising 15 nm thick HfO$_2$ grown at 90$^\circ$C using atomic layer deposition (ALD), was then deposited to galvanically isolate the ohmic Ti/Au layer from the rest of the device. This was followed by the deposition of Ti/Au layers (5~nm/300~nm) for the inner ohmic contact. Next, a global layer of 15~nm thick HfO$_2$ was deposited over the entire sample to serve as the gate dielectric. Gates were defined using electron beam lithography followed by e-beam evaporation of Ti/Au layers of thickness (5~nm/20~nm) for finer structures and (5~nm/350~nm) for the bonding pads.

\vspace{10 pt}

{\bf Three-terminal electrical transport measurements:} Electrical transport measurements were performed in an Oxford Triton dilution refrigerator at a base temperature of $\sim$20 mK using standard AC lock-in techniques. The superconducting loop was grounded by connecting it to the fridge ground through a $\sim$0.9~k$\Omega$ line. The top and the bottom ohmic contacts were connected to low-impedance current-to-voltage converters, through $\sim$1.5--2.0~k$\Omega$ line resistances. The grounding pin of each current-to-voltage converter was biased through a combination of AC+DC voltages $V_{\rm t(b)}+V_{\rm T(B)}$, at the top and bottom ends respectively. The AC biases, $V_{\rm t}$ and $V_{\rm b}$, were generated by two lock-in amplifiers, with the same excitation amplitude (3~$\mu$V), but different frequencies, $f_{\rm t}=31.5$~Hz and $f_{\rm b}=77.5$~Hz respectively. The DC biases $V_{\rm T(B)}$ were generated using two low-noise DC voltage sources. The voltage output of each current-to-voltage converter was measured using two lock-in amplifiers, operating at frequencies $f_{\rm t}$ and $f_{\rm b}$, requiring a total of 4 lock-in amplifiers for each element of the 2$\times$2 conductance matrix.

\begin{figure*}[htb]
\includegraphics[width=1\textwidth]{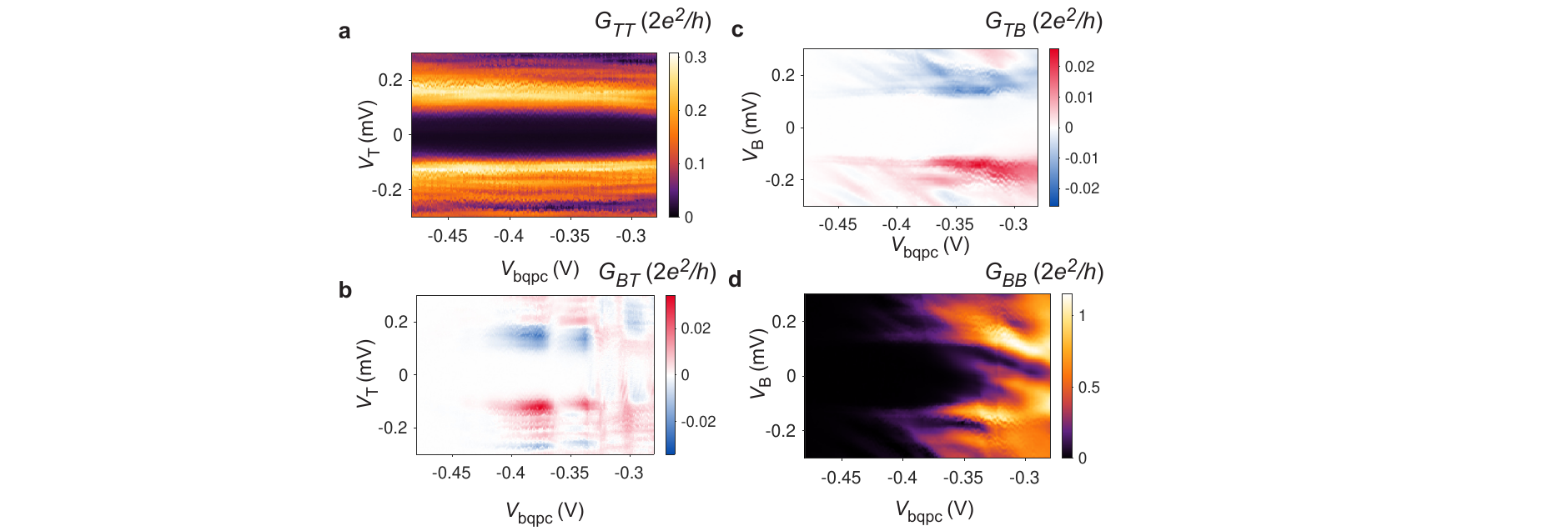} 
\caption{\label{suppfigS1} {\bf Conductance matrix as a function of bottom QPC voltages in Device 1}~ (a)-(b) Conductance matrix measured as function of bottom QPC gate voltage $V_{\rm bqpc}$ and top end source-drain bias $V_{\rm T}$, with $V_{\rm B}=$~0 and $V_{\rm tqpc}=-0.37$~V. (a) $G_{\rm TT}$ shows a hard superconducting gap at the top end as a function of $V_{\rm T}$, but does not respond to $V_{\rm bqpc}$. (b) $G_{\rm BT}$ exhibits anti-symmetry and is pinched off by $V_{\rm bqpc}$. (c)-(d) Conductance matrix measured as function of bottom QPC gate voltage, $V_{\rm bqpc}$ and bottom end source-drain bias, $V_{\rm B}$, with $V_{\rm T}=0$ and $V_{\rm tqpc}=-0.37$~V. (c) $G_{\rm TB}$, measured at the top end, is pinched off by $V_{\rm bqpc}$ at the bottom end, reflecting nonlocality. (d) $G_{\rm BB}$ shows a hard superconducting gap and is pinched off by $V_{\rm bqpc}$.}
\end{figure*}

\begin{figure*}[htb]
\includegraphics[width=1\textwidth]{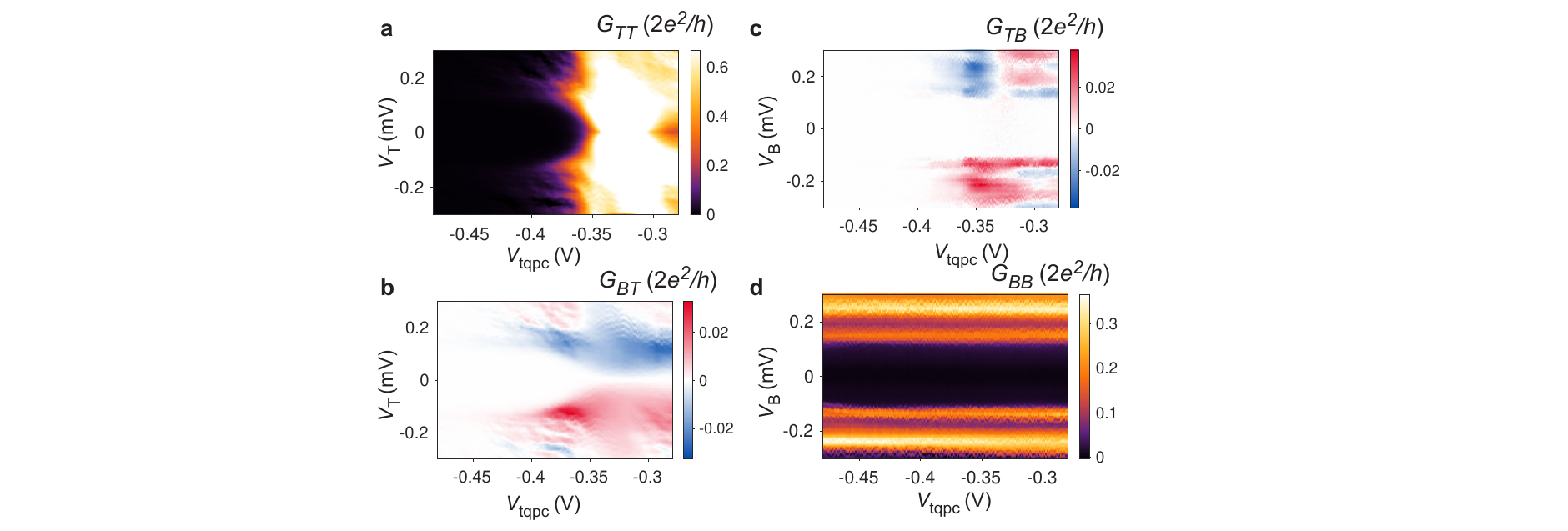} 
\caption{\label{suppfigS2} {\bf Conductance matrix as a function of top QPC voltages in Device 1.}~(a)-(b) Conductance matrix measured as function of top QPC gate voltage, $V_{\rm tqpc}$ and top end source-drain bias, $V_{\rm T}$, with $V_{\rm B}=0$ and $V_{\rm bqpc}=-0.38$~V. (a) $G_{\rm TT}$ shows a hard-superconducting gap and is pinched off by $V_{\rm tqpc}$. (b) $G_{\rm BT}$, measured at the bottom end, is pinched off by $V_{\rm tqpc}$ at the top end, reflecting nonlocality. (c)-(d) Conductance matrix measured as function of top QPC gate voltage $V_{\rm tqpc}$ and bottom end source-drain bias $V_{\rm B}$, with $V_{\rm T}=$~0 and $V_{\rm bqpc}=-0.38$~V. (c) $G_{\rm TB}$, exhibits antisymmetry with respect to zero-bias and is pinched off by $V_{\rm tqpc}$. (d) $G_{\rm BB}$ shows a hard superconducting gap at the bottom end as a function of $V_{\rm B}$, but does not respond to $V_{\rm tqpc}$.}

\end{figure*}

We find experimentally that the conductance $G_{ij}$ depends strongly on $V_j$ but only weakly on $V_i$ (unless $i=j$), as shown in Figs.~\ref{suppfigS1} and ~\ref{suppfigS2}. This is expected theoretically~\cite{DanonNonlocal}. Therefore, for measurement of $G_{\rm TT}$ and $G_{\rm BT}$, we sweep the DC bias $V_{\rm T}$ and set $V_{\rm B}=0$, and for measurement of $G_{\rm BB}$ and $G_{\rm TB}$, we sweep the DC bias $V_{\rm B}$ and set $V_{\rm T}=0$. 

\vspace{10 pt}

{\bf Voltage-divider effects:}
In the three-terminal geometry, measurement circuit effects arising from finite line impedances can lead to spurious voltage-divider effects as discussed in Ref.~\cite{martinez2021}. DC and AC voltage drops across line-resistances connected to the top ($R_{\rm T}$), bottom ($R_{\rm B}$), and ground ($R_{\rm G}$) terminals of the devices produce contributions to the local and nonlocal conductance signals, and can be corrected using the method outlined in Ref.~\cite{martinez2021}.

Briefly, the $2\times2$ conductance matrix $G_{ij}$ ($i$,~$j$ = T,~B) measured as a function of voltages $V_{\rm T,B}$ applied outside the cryostat, is converted to a corrected conductance matrix $G^{\prime}_{ij}$ obtained as a function of $U_{\rm T,B}$~, the voltages estimated to be applied at the top and bottom terminals of the device inside the cryostat, using the relation ${\bf G}^{\prime}= {\bf G} [\bf{I}_2 - {\bf Z}{\bf G}]^{-1}$. Here $\bf Z=\begin{bmatrix}
R_{\rm T} & R_{\rm T} + R_{\rm G}\\
R_{\rm B} + R_{\rm G} & R_{\rm B}\\
\end{bmatrix}$ is the impedance matrix. 

As an example, we depict in Fig.~\ref{suppfigS12} the corrected conductance matrix for data shown in Fig.~\ref{fig02}. The line resistances were measured as $R_{\rm T} \simeq 2~{\rm k} \Omega$, $R_{\rm B} \simeq 2~{\rm k}\Omega$, and $R_{\rm G} \simeq 0.9~{\rm k}\Omega$. As shown in Fig.~\ref{suppfigS12}(b), the voltage-divider corrections do not produce qualitative changes to the uncorrected data. 

\vspace{10 pt}

{\bf Phase-biasing and magnetic field alignment:}  Magnetic field to the sample is applied using a three-axis ($B_x$,~$B_y$,~$B_z$)=(1~T,~1~T,~6~T) vector rotate magnet. The sample is oriented with respect to the vector magnet such that $B_x$ is nominally parallel to $B_\perp$ [Fig.~\ref{fig01}(c)] and $B_z$ is nominally parallel to $B_\parallel$ [Fig.~\ref{fig01}(a)]. However, because of sample misalignment, $B_z$ has a small but finite contribution to $B_\perp$, which controls the flux through the phase-biasing loop. For a finite value of $B_z$, is is therefore necessary to identify the value of $B_x$  that leads to lines of constant flux. At zero $B_z$, the value of $B_x$ at which the superconducting gap is maximised corresponds to zero and multiples of $\Phi_0$, while distinct phase-slips appear at odd multiples of $\Phi_0/2$. This allows calibration of the flux through the device at zero $B_z$. At finite $B_z$, the superconducting gap acquires a phase-asymmetric dispersion, and the maxima of the gap cannot be used to track lines of constant flux. Instead, we use the phase-slips, which become more prominent at high values of $B_z$, to identify lines of constant flux through the device. The values of $B_x$ required for this compensation are typically linear as a function of $B_z$.

\vspace{10 pt}

{\bf Numerical simulations:} To model our device, we use an extension of the Hamiltonian proposed in Refs.~\cite{Pientka,HellFlensberg} utilizing a two-layer structure to account for finite thickness and include orbital effects, as described in more detail in Ref.~\cite{GapReopening}. Here we briefly describe the essentials of the model.
The rectangular device contains three parts: normal region in the middle with width $w_n$, and superconducting regions of width $w$ on each side. In the Nambu basis $(\psi_{\uparrow},\psi_{\downarrow},\psi_{\downarrow}^\dagger,-\psi_{\uparrow}^\dagger)$, the Bogoliubov--de Gennes Hamiltonian is given by
\begin{equation}\label{eq:Hamiltonian}
\begin{aligned}
    H&=\left[-\frac{\partial^2_x+\partial^2_y}{2m^*}-t_{\perp}\nu_x-\mu(y)+i\alpha(z)\left(\partial_x\sigma_y-\partial_y\sigma_x\right)\right]\tau_z \\
    &+\frac{g(y)\mu_{\rm B} B_\parallel}{2}\sigma_x+\Delta(y,z)\tau^++\Delta^*(y,z)\tau^-.
\end{aligned}
\end{equation}
Here $\sigma,\tau,\nu$ are Pauli matrices acting in spin, electron-hole, and layer basis, respectively. $m^*$ is the effective mass of electrons in the semiconductor, $\alpha(z)$ is the layer-dependent Rashba spin-orbit coupling strength, $t_{\perp}$ is the inter-layer hopping amplitude, $B_\parallel$ is the magnetic field applied along the junction, and $\mu_{\rm B}$ is the Bohr magneton. The g-factor $g(y)$ and the chemical potential $\mu(y)$ are $g_{\rm N},\mu_{\rm N}$ in the normal region and $g_{\rm S},\mu_{\rm S}$ in the superconducting regions.
$\Delta(y)$ is the superconducting pairing potential which is non-zero only in the superconducting region and includes a phase difference $\phi$:
\begin{equation}
    \Delta(y)=\left\{ \begin{array}{ll}
\Delta e^{i\phi/2} & \, \frac{w_n}{2}<y<w+\frac{w_n}{2}\\
0 & \, |y|<\frac{w_n}{2}\\
\Delta e^{-i\phi/2} & \, -w-\frac{w_n}{2} < y < -\frac{w_n}{2}.
\end{array} \right.
\end{equation}
Orbital effects due to the in-plane magnetic field are accounted for by adding a complex phase factor to the inter-layer hopping amplitude $t_{\perp}$, corresponding to the vector potential $\vec{A}=B_{\parallel}y\hat{z}$~\cite{peierls_zur_1933}.
The parallel magnetic field also induces linear phase growth along the junction's cross section~\cite{tinkham_introduction_2004}, which is modeled as an additional modulation $\propto B_{\parallel}yz$ to the superconducting phase.

For the purpose of numerical simulations, we discretize the Hamiltonian to a tight-binding model on a square lattice of spacing $a=10$~nm. Simulations are performed with the following parameters: $m^* = 0.026\,m_e$, $\Delta=140~\mu$eV, $t_{\perp}=10$~meV, $l=10~\mu$m, $w_n=100$~nm, $w = 200$~nm, $w_z=10$~nm, $\mu_{\rm SC}=3.6$~meV, $\mu_{\rm N}=3.3$~meV. The g-factors are taken to be $g_{\rm N}=8$ and $g_{\rm S}=4$, and the spin-orbit coupling constants are $\alpha(0) = 15$~meV~nm, $\alpha(1)=-\alpha(0)/4$.
Transport simulations were performed using the Kwant software package~\cite{groth_kwant_2014}. We attach two normal leads to the top and bottom of the device, and add tunneling barriers of $25~{\rm meV}$ between the leads and the device. We then numerically solve the scattering problem and extract the conductance matrix.
To account for thermal broadening, we perform a convolution of the conductance with the derivative of the Fermi-Dirac function. The temperature we used in these calculations is 50~mK.

\clearpage

\begin{figure}[h]
\includegraphics[width=0.5\textwidth]{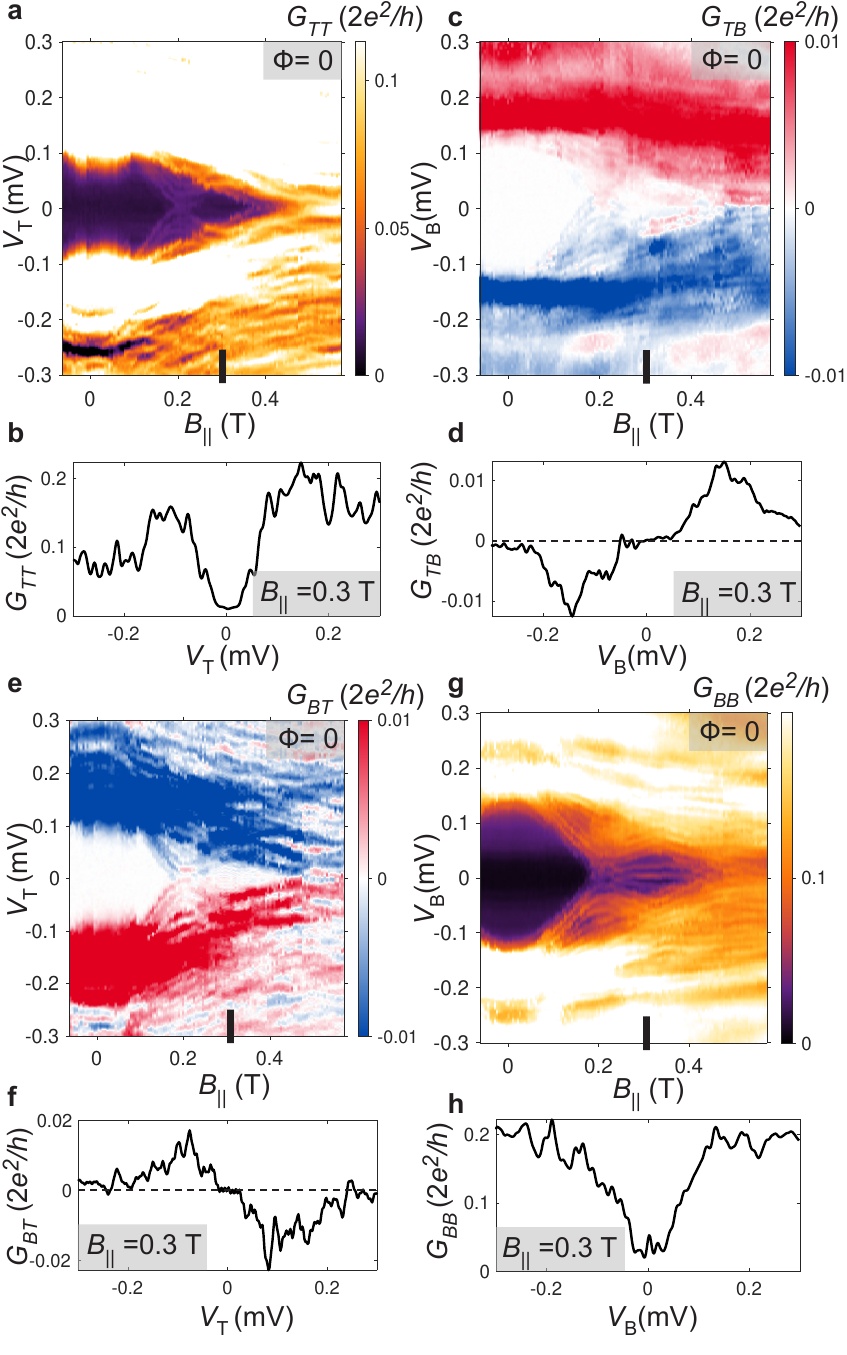}
\caption{\label{suppfigS3} {\bf Conductance matrix as a function of in-plane magnetic field in a regime supporting ZBCP only at bottom end in Device 1.}~(a)~Local differential conductance $G_{\rm TT}$ and (e)~Nonlocal differential conductance $G_{\rm BT}$ measured as a function of $V_{\rm T}$ and $B_{\parallel}$. (c)~Nonlocal differential conductance $G_{\rm TB}$ and (g)~local differential conductance $G_{\rm BB}$ measured as a function of $V_{\rm B}$ and $B_\parallel$. The phase bias is set to $\Phi=0$. Line-cuts at $B_\parallel=0.3$~T where (b)~$G_{\rm TT}$ is gapped. (h)~$G_{\rm BB}$ shows a ZBCP. (d)~$G_{\rm TB}$ and (f)~$G_{\rm BT}$ are strongly antisymmetric at high DC biases, and zero in a finite range of DC bias around zero.}
\end{figure}

\begin{figure*}[htb]
\includegraphics[width=1\textwidth]{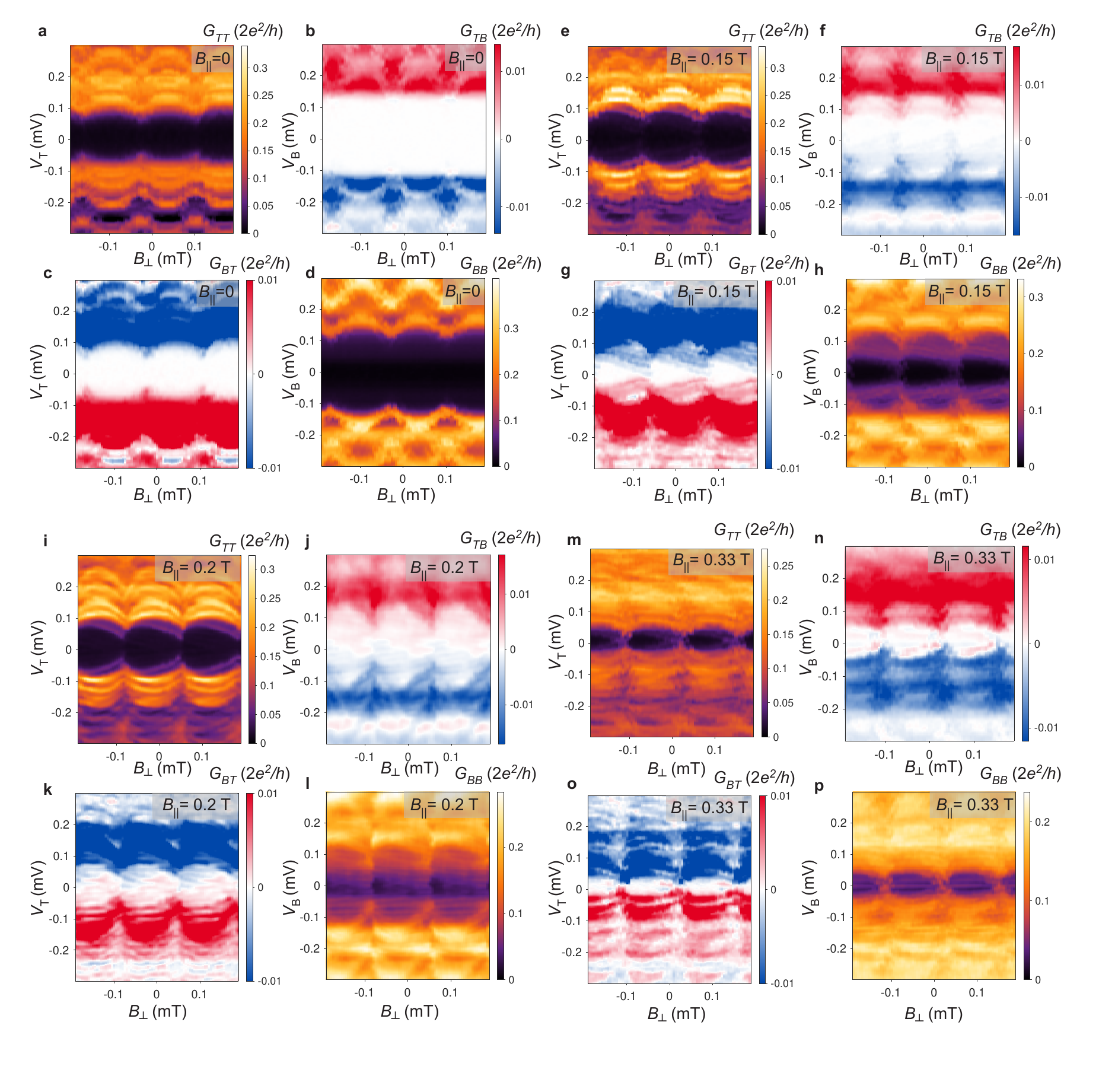} 
\caption{\label{suppfigS4} {\bf Phase dependence of the conductance matrix in Device 1.}~Conductance matrix measured as a function of out-of-plane magnetic field at different values of in-plane magnetic field. (a)--(d) $B_\parallel$=0. All four conductance matrix elements exhibit periodic modulation of the superconducting gap. (e)--(h) $B_\parallel$=0.15~T. Subgap states are lowered in energy. These states create a phase-asymmetric modulation of the gap within each flux lobe. (i)--(l) $B_\parallel$=0.2~T. The superconducting gap is closed at all values of $\Phi$, the closure being most evident in $G_{\rm BT}$. (m)--(p) $B_\parallel$=0.33~T. The superconducting gap reopens with a phase independent ZBCP visible in $G_{\rm BB}$. Other conductance matrix elements show a phase-modulated superconducting gap.}
\end{figure*}

\begin{figure*}[htb]
\includegraphics[width=1\textwidth]{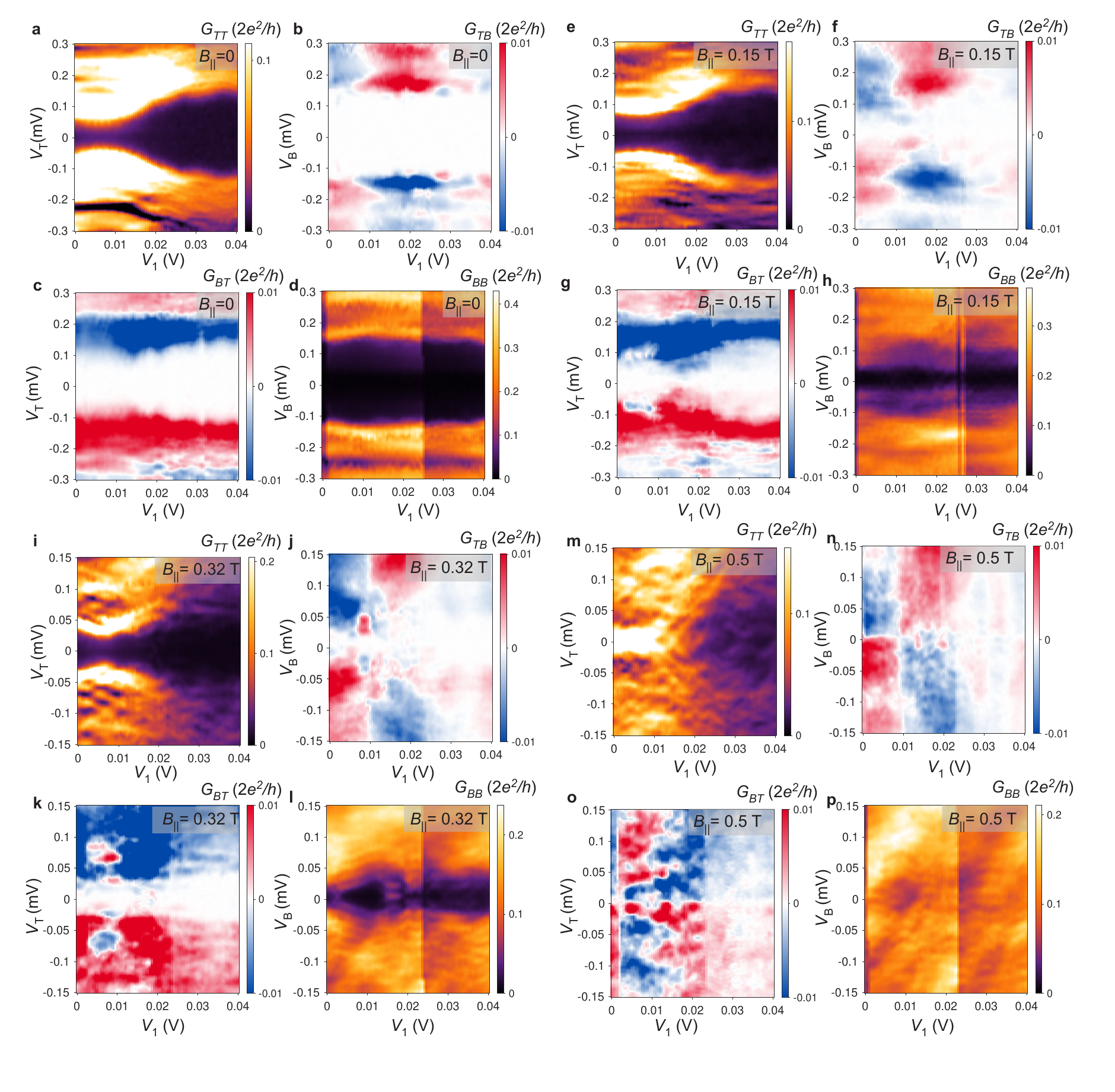} 
\caption{\label{suppfigS5}{\bf Top gate voltage dependence of the conductance matrix in Device 1.}~Conductance matrix measured as a function of top gate voltage $V_1$ at different values of in-plane magnetic field. We set $\Phi=0$. (a)--(d) $B_\parallel=0$. All four conductance matrix elements show a maximum superconducting gap of size $\sim 150~\mu$V. $G_{\rm TT}$ shows a suppression of the gap near $V_1\sim$~5 mV, which is not observed in the other conductance matrix elements. We interpret this as a local feature at the top end. (e)--(h) $B_\parallel=0.15$~T. Subgap states are lowered in energy, but the superconducting gap remains finite as a function of $V_1$. (i)--(l) $B_\parallel=0.32$~T. Finite superconducting gap measured in the reopened state, particularly visible in the nonlocal conductance matrix elements. $G_{\rm BB}$ shows a ZBCP centered around $V_1=17$~mV. (m)--(p) $B_\parallel=0.5$~T. Final closure of the gap, visible in all four conductance matrix elements. Note that the nonlocal conductance signal is strongly antisymmetric and concentrated around zero bias.}
\end{figure*}

\begin{figure*}[htb]
\includegraphics[width=1\textwidth]{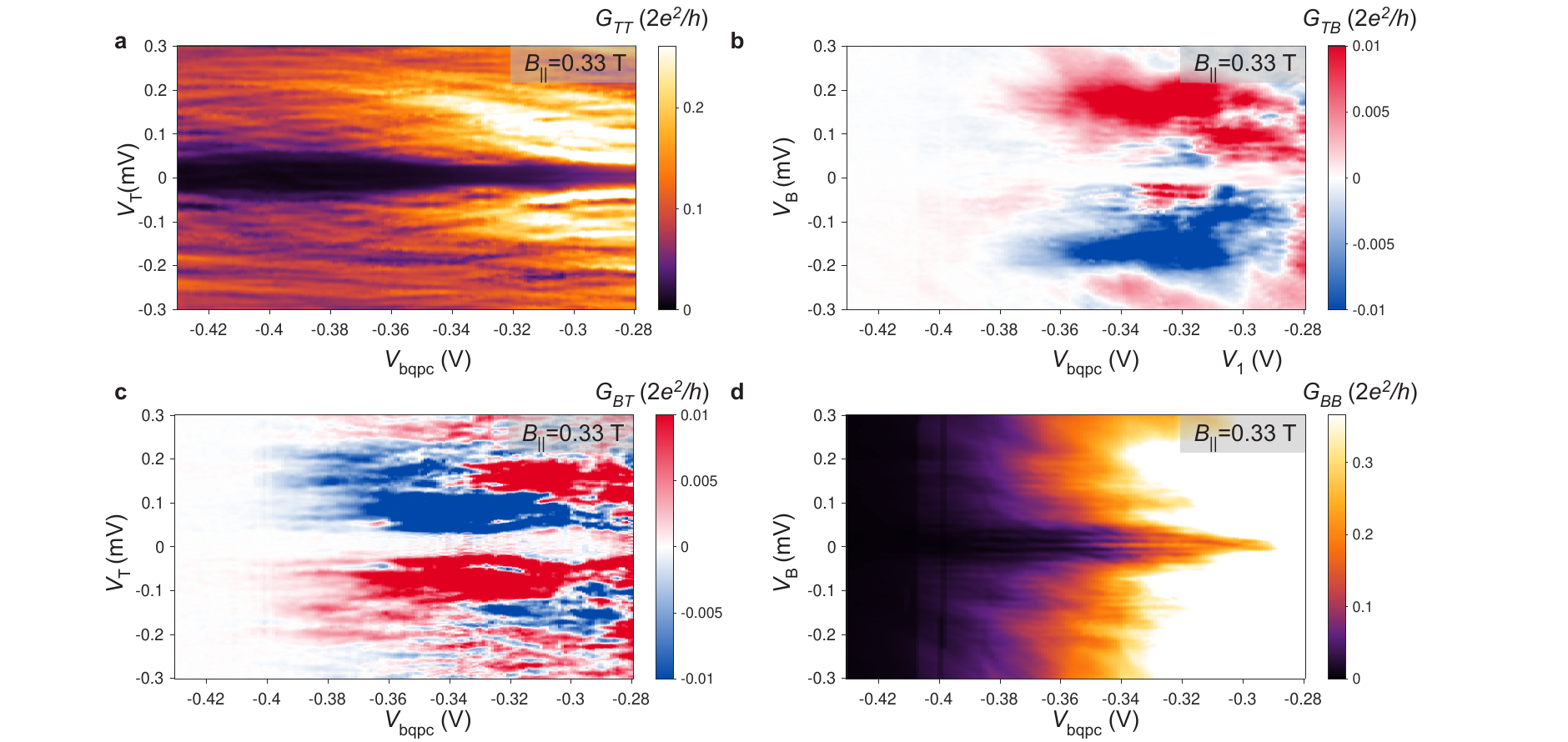} 
\caption{\label{suppfigS6} {\bf Conductance matrix as a function of bottom QPC pinch-off in Device 1.}~Conductance matrix measured as a function of bottom QPC gate voltage $V_{\rm bqpc}$ at $V_1$=~17~mV , $B_\parallel=$~0.33~T and $\Phi=0$. (a) $G_{\rm TT}$ shows a superconducting gap without subgap structure. The gap is modulated weakly by $V_{\rm bqpc}$. (b) $G_{\rm TB}$ is strongly antisymmetric and shows a superconducting gap. The signal is diminished and ultimately pinched off by $V_{\rm bqpc}$. (c) $G_{\rm BT}$ is strongly antisymmetric and shows a superconducting gap. The signal is diminished and ultimately pinched off by $V_{\rm bqpc}$. (d) $G_{\rm BB}$ shows a sharp ZBCP that becomes visible at $V_{\rm bqpc} \sim -0.31$~V and persists until pinch-off as a function of $V_{\rm bqpc}$.}
\end{figure*}

\begin{figure*}[htb]
\includegraphics[width=1\textwidth]{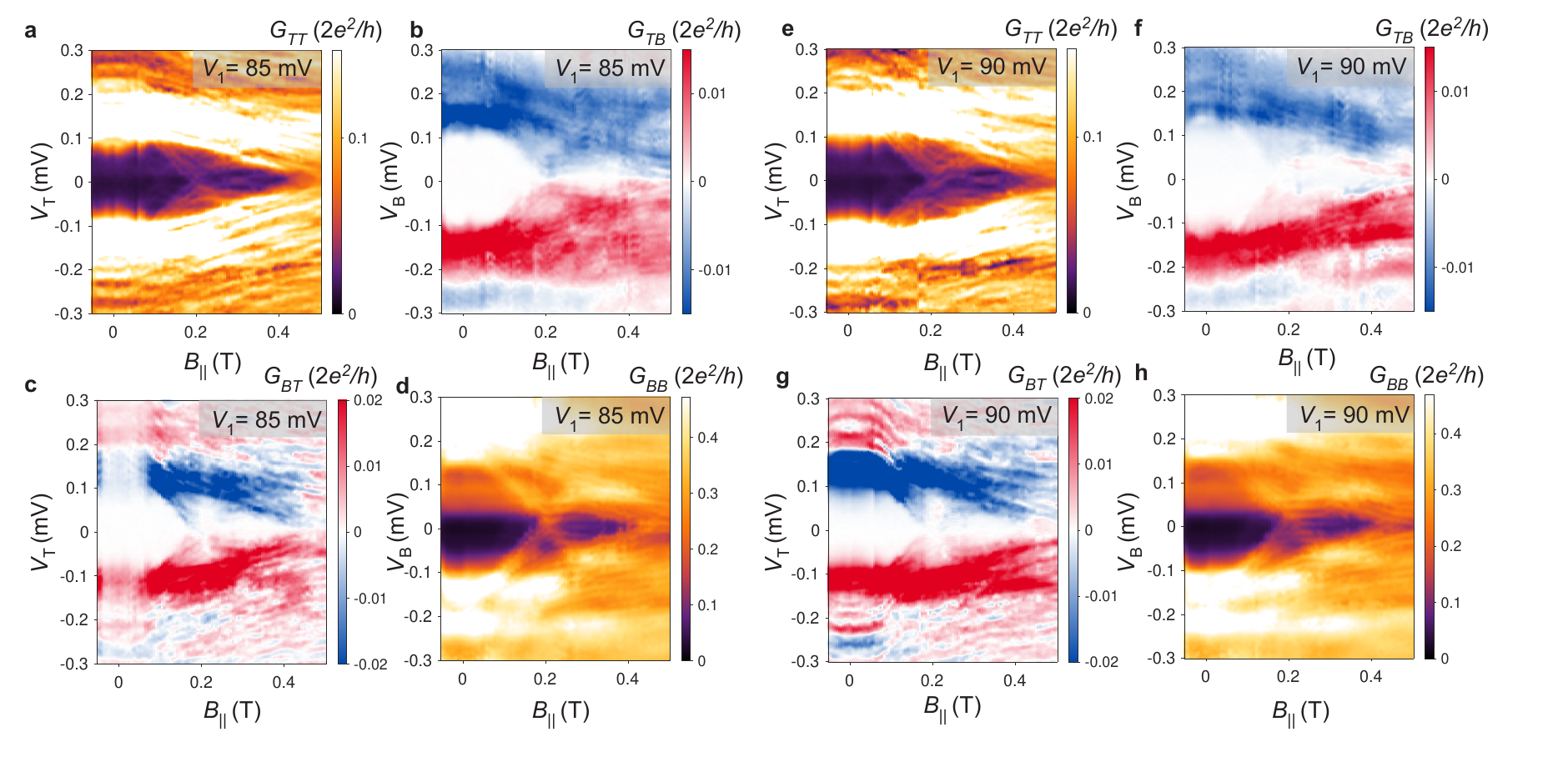}
\caption{\label{suppfigS7} {\bf Conductance matrix as a function of in-plane magnetic field in regimes supporting ZBCPs at both ends in Device 1}~ (a)--(d) $V_1~$=~85~mV and (e)--(h) $V_1~$=~90~mV.  Both local conductances show a gap-reopening transition at $B_\parallel=$~0.2~T followed by ZBCPs. The nonlocal conductances are strongly antisymmetric with respect to bias, undergo a gap-reopening transition at $B_\parallel=$~0.2~T, and show a reopened gap that is devoid of sub-gap structure.}
\end{figure*}

\begin{figure*}[htb]
\includegraphics[width=1\textwidth]{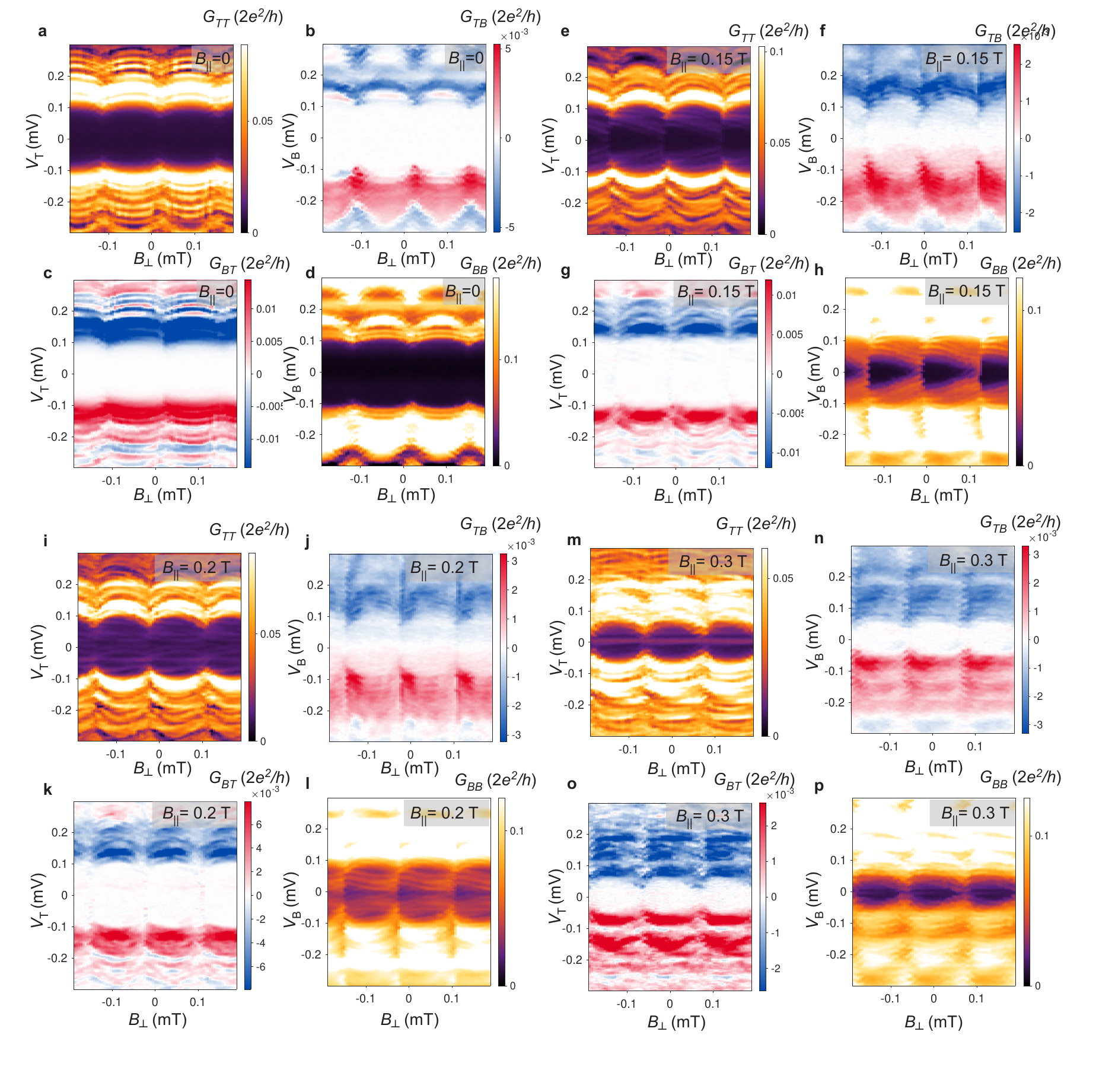}
\caption{\label{suppfigS8} {\bf Conductance matrix measured as a function of out-of-plane magnetic field at different values of in-plane magnetic field, at $V_1~$=~85~mV in Device 1.}~(a)--(d) $B_\parallel=0$. All four conductance matrix elements exhibit periodic modulation of the superconducting gap. (e)--(h) $B_\parallel=0.15$~T. Subgap states are lowered in energy. These states create a phase-asymmetric modulation of the gap within each flux lobe. (i)--(l) $B_\parallel$=0.2~T. The superconducting gap is closed at all values of $\Phi$. (m)--(p) $B_\parallel=0.3$~T. The superconducting gap reopens with a phase independent ZBCP visible in both $G_{\rm TT}$ and $G_{\rm BB}$. The nonlocal conductance matrix elements show a phase-modulated superconducting gap, with no sub-gap structure.}
\end{figure*}

\begin{figure*}[htb]
\includegraphics[width=1\textwidth]{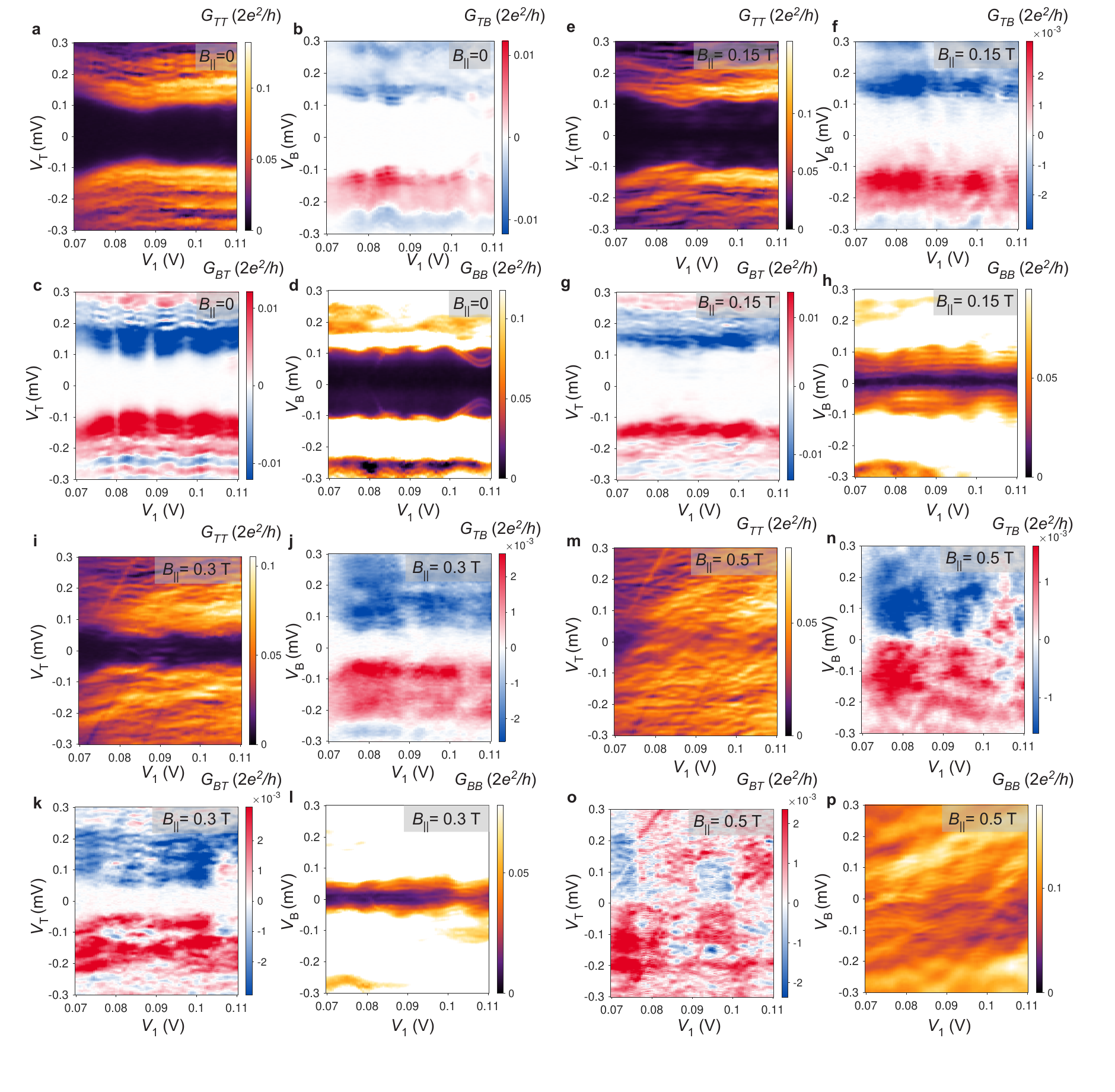}
\caption{\label{suppfigS9} {\bf Conductance matrix measured as a function of top gate voltage $V_1$ at different values of in-plane magnetic field in Device 1}~(a)--(d) $B_\parallel=0$. All four conductance matrix elements show a superconducting gap of $\sim 150~\mu$V, with some dependence on $V_1$. (e)--(h) $B_\parallel=0.15$~T. Sub-gap states are lowered in energy, but the superconducting gap remains finite as a function of $V_1$. (i)--(l) $B_\parallel=0.3$~T. Finite superconducting gap measured in the reopened state, particularly visible in the nonlocal conductance matrix elements. Both $G_{\rm TT}$ and $G_{\rm BB}$ show ZBCPs existing in voltage range $V_1=$~0.08 -- 0.1~V. (m)--(p) $B_\parallel=0.5$~T. Final closure of the gap, visible in all four conductance matrix elements. $\Phi=0$ for all plots.}
\end{figure*}

\begin{figure*}[htb]
\includegraphics[width=1\textwidth]{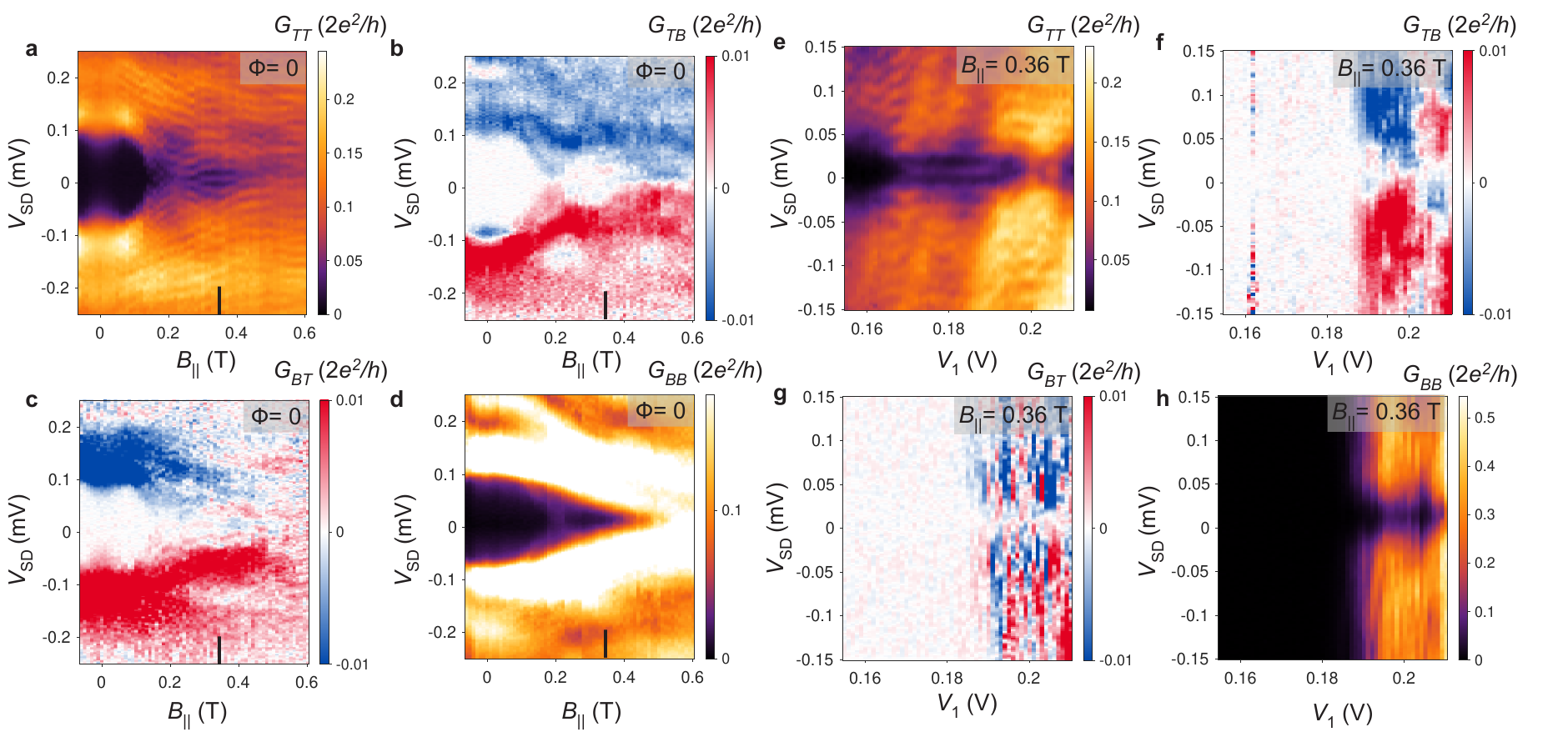}
\caption{\label{suppfigS10}{\bf Conductance matrix in Device 2}~(a)--(d) In-plane magnetic field dependence of the conductance matrix at $V_1~=185$~mV. Both local conductances show a gap-reopening transition at $B_\parallel=$~0.2~T, with a ZBCP appearing only at the top end. The nonlocal conductances are strongly antisymmetric with respect to bias, undergo a gap-reopening transition at $B_\parallel=0.2$~T. (e)--(h) Conductance matrix at $B_\parallel=0.3$~T, $\Phi=0$ measured as a function of $V_1$. $G_{\rm TT}$ shows a stable ZBCP, whereas $G_{\rm BB}$ remains pinched off for $V_1<0.19$~V. Both $G_{\rm TB}$ and $G_{\rm BT}$ exhibit finite sub-gap nonlocal conductance for $V_1= 0.19-0.21$~V}
\end{figure*}

\begin{figure*}[htb]
\includegraphics[width=1\textwidth]{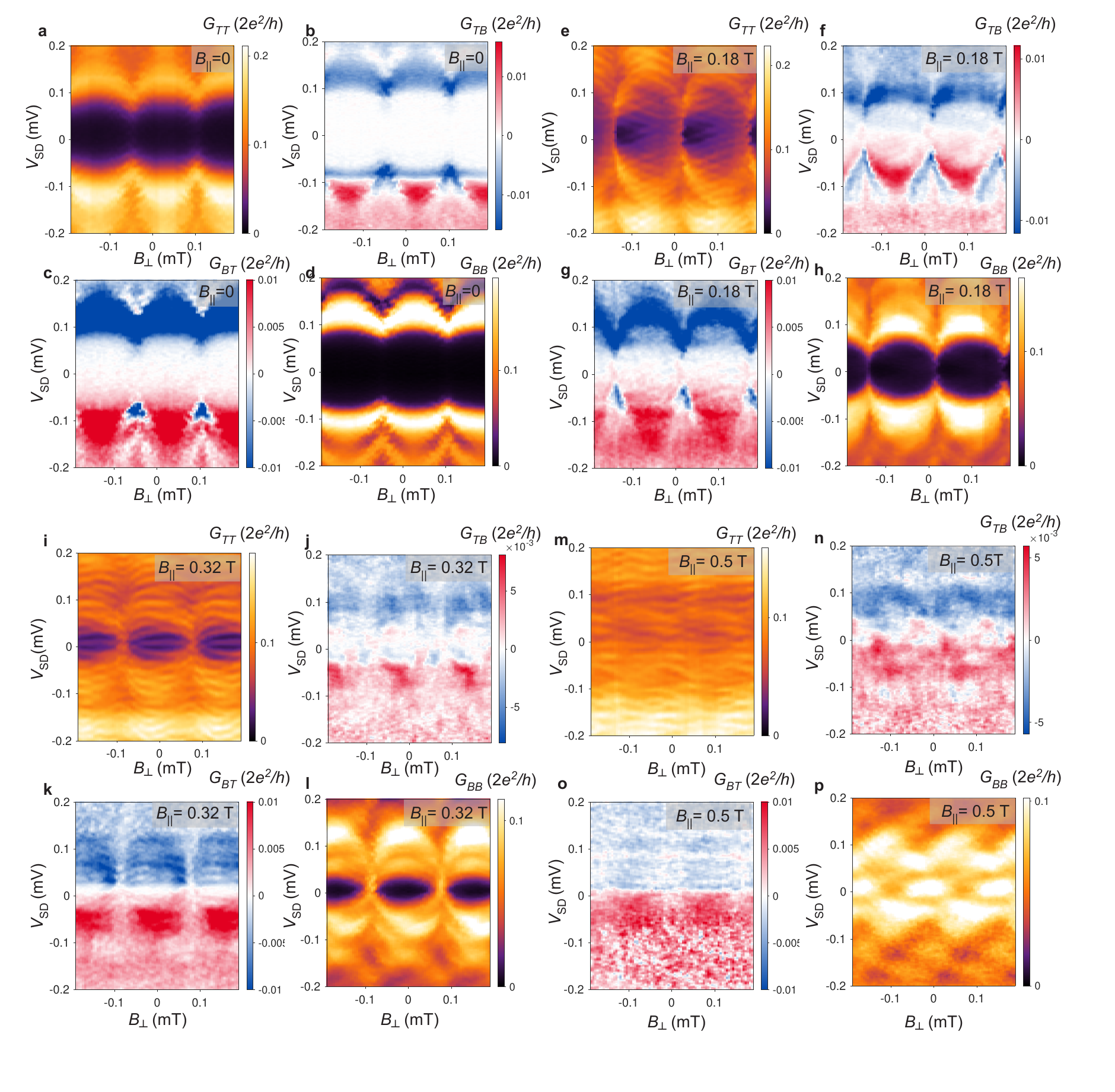}
\caption{\label{suppfigS11}{\bf Phase dependence of the conductance matrix in Device 2.} Conductance matrix measured as a function of out-of-plane magnetic field at different values of in-plane magnetic field, at $V_1~= 0.185$~V. (a)--(d) $B_\parallel=0$. All four conductance matrix elements exhibit periodic modulation of the superconducting gap. (e)--(h) $B_\parallel=0.18$~T. Subgap states are lowered in energy. These states create a phase-asymmetric modulation of the gap within each flux lobe, particularly visible in $G_{\rm TT}$.   (i)--(l) $B_\parallel=0.3$~T. The superconducting gap reopens with a phase independent ZBCP visible in $G_{\rm TT}$. The nonlocal conductance matrix elements show a phase-modulated superconducting gap. (m)--(p) $B_\parallel=0.5$~T. The superconducting gap is closed at all values of $\Phi$.}
\end{figure*}

\begin{figure*}[htb]
\includegraphics[width=1\textwidth]{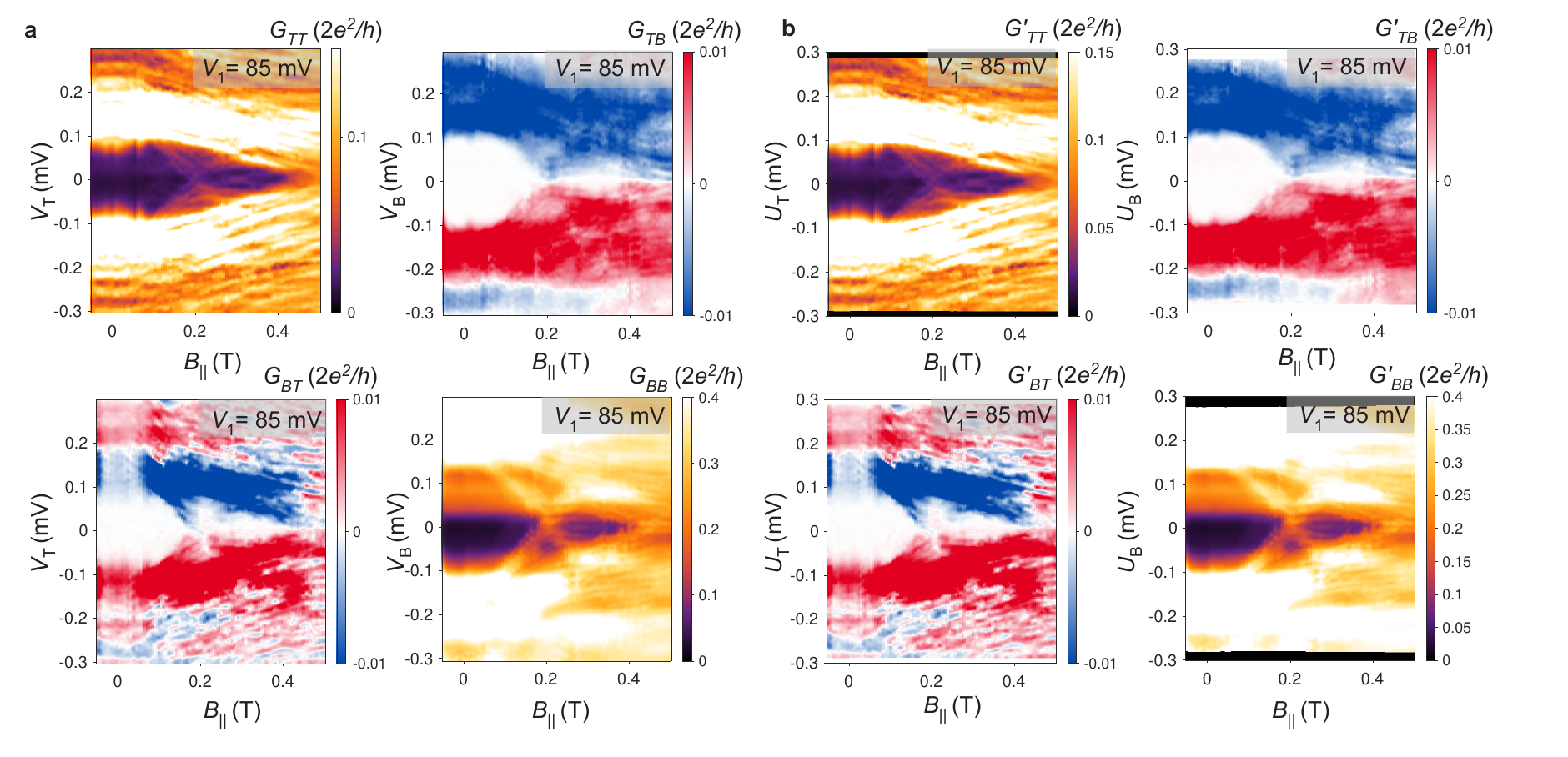}
\caption{\label{suppfigS12} {\bf Voltage-divider effects in Device 1.} (a)~Conductance matrix, uncorrected for voltage-divider effects, as a function of in-plane magnetic field at $V_1~=~85$~mV (same data as Fig.~\ref{fig02}). (b)~Conductance matrix corrected for voltage-divider effects, as a function of in-plane magnetic field at $V_1~=~85$~mV. $U_{\rm T}$ and $U_{\rm B}$ represent the estimated voltage biases appearing at the top and bottom terminals of the device, after accounting for voltage drops across line resistances $R_{\rm T} \simeq 2~{\rm k} \Omega$ connected to the top terminal, $R_{\rm B} \simeq 2~{\rm k}\Omega$ connected to the bottom terminal, and $R_{\rm G} \simeq 0.9~{\rm k}\Omega$ connected to the ground terminal. The conductance corrections arising from voltage-divider effects do not result in any qualitative changes.} 
\end{figure*}

\end{document}